# Shear-stress-constrained superconductivity in Ruddlesden-Popper nickelates

Liling Sun[1,2], Shu Cai[1], Jinyu Zhao[1], Qi Wu[2], Yang Ding[1], Tao Xiang[2], and Ho-kwang Mao[1,3]

[1]*Center for High Pressure Science & Technology Advanced Research, Beijing 100193, China*
[2]*Institute of Physics, Chinese Academy of Sciences, Beijing 100190, China*

[3] *Shanghai Key Laboratory of MFree, Shanghai Advanced Research in Physical Sciences, Shanghai, 200003, China*

Ruddlesden-Popper nickelates exhibit superconductivity under pressure in bulk crystals and under epitaxial constraint in thin films, while remaining highly sensitive to sample quality, oxygen content, defects, and stress conditions. We propose that the metastable RP lattice becomes superconducting only when the local constrained deformation of the Ni-O framework falls within a bounded shear-strain window. This deformation controls octahedral rotations, the interlayer Ni-O-Ni bond angle, and coupling between Ni $dz^2$ and $dx^2$-$y^2$ orbitals. This shear-stress-constrained superconductivity scenario unifies the understanding of the pressure threshold, reversibility, spatial inhomogeneity, pressure medium dependence, film-substrate sensitivity, and reproducibility.

High-temperature superconductivity in strongly correlated 3*d* transition-metal oxides with Ruddlesden-Popper (RP) structure is typically achieved by stabilizing a specific constraint lattice and electronic state. External pressure, epitaxial strain, chemical substitution, oxygen stoichiometry and intergrowth control bond angles, bond lengths, orbital occupation, and carrier distribution. In the oxides, these variables are strongly entangled. The central challenge is therefore to identify the local constraint that selects the superconducting ground state, rather than simply to assign superconductivity to pressure, strain, or crystallographic symmetry alone.

RP nickelates provide a particularly sharp example of this general problem. High-$T_c$ superconducting signatures have been reported in both pressure-constrained bulk samples[1-5] and substrate-constrained thin films[6-12], which have become an active subject in high-$T_c$ superconductivity research[13-18]. Yet these systems exhibit strong sample dependence, spatially inhomogeneous phase distributions [19-21], sensitivity to oxygen content[19,22,23] and intergrowth phases[24-26], and apparently different pressure or strain requirements in bulk and film geometries. These observed phenomena demand a unified understanding of how superconductivity emerges and how it is stabilized by external or internal constraints.

For compressed bulk nickelates, such a framework must account for why pressures above ~14 GPa are needed [1,27], why the shear-stress distribution is inhomogeneous [20], why an ambient-pressure tetragonal phase is not superconducting even after compression[28], why chemical doping or hybrid structures can enhance superconductivity[3,4,7,8,11,26,29], why superconductivity is inhomogeneous or filamentary-like [19,20,21], why superconductivity is sensitive to the pressure-transmitting medium[2,4,29], and why the transition is reversible[19]. For epitaxial films, it must explain why further compression can raise $T_c$ [30,31], why superconductivity is sensitive to substrate quality [6.7], and why a sufficient shrinkage or constraint is required. These phenomena involve the constrained lattice symmetries and stress distributions that differ between bulk and film RP nickelates. Here, we propose that these phenomena can be understood in a unified way by a shear-stress-constrained superconductivity (SSCS) scenario.

## 1. Experimental basis for SSCS

Key experimental evidence for SSCS comes from local measurements on compressed $La_3Ni_2O_7$ that combine nitrogen-vacancy quantum sensing of diamagnetism with resistance measurements and stress mapping in a diamond-anvil cell[20]. These experiments directly provide normal stress, shear stress, and stoichiometry in compressed $La_3Ni_2O_7$, align with the superconducting response, and conclude that superconductivity disappears when the shear stress exceeds ~2 GPa. A closer reading of the reported data further suggests that superconductivity is also weakened when the shear stress approaches zero. Thus, shear stress is not simply a detrimental by-product of nonhydrostatic pressure, instead, superconductivity appears to require a finite, optimal shear-stress window. This indicates that a scalar pressure value is insufficient to identify the origin of the superconducting state; the local stress tensor, stoichiometry, and structural constraint must be considered together.

This observation requires careful distinction between stress and the microscopic distortion that enters the electronic Hamiltonian. Shear stress is the experimentally accessible variable, while the electronic system responds directly to the resulting local strain, bond-angle change, bond-length change, and orbital reconstruction. SSCS should therefore be read as a bounded local constraint. A hydrostatic pressure or epitaxial substrate may help stabilize the metastable bond geometry, but excessive shear broadens the strain distribution, promotes defects or domain imbalance, thus destroys the coherence needed for superconductivity.

## 2. Metastable RP lattice and exotic shear stress

The RP lattice is thermodynamically metastable and kinetically stabilized. Its layered structure contains an intrinsic mismatch between perovskite blocks and rock-salt layers, which results in internal stress, octahedral tilting, and stacking faults, as intrinsic structural characteristics. This metastability explains why RP nickelates are unusually sensitive to oxygen stoichiometry, external stress, epitaxial strain, chemical substitution, and structural defects. Small changes in any of these variables can modify the coupled lattice, orbital, and charge degrees of freedom.

In compressed $La_3Ni_2O_7$, superconductivity appears when the ambient-pressure orthorhombic structure is driven toward a high-pressure tetragonal-like structure. During this transformation the $NiO_6$ octahedra rotates, and the interlayer Ni-O-Ni bond angle approaches the nearly linear configuration favorable for effective interlayer hopping (see Fig.1a). This rotation tunes bond angles, bond lengths, orbital polarization, and carrier distribution simultaneously, inducing a shear stress that becomes a key factor in altering the electronic structures. The essential point in the SSCS scenario is that this rotation does not merely produce a higher-symmetry lattice; it stores elastic shear strain in the constrained Ni-O-Ni bonds. Superconductivity is then associated with a metastable constrained state in which both bond angle and bond length are tuned into a favorable range for the relevant orbital and electronic structure to develop superconductivity. This distinction explains why an ambient-pressure tetragonal phase cannot become superconducting under pressure[28] (see Fig.1b), therefore tetragonal symmetry and *in-plane* compression are not sufficient if the sample lacks the appropriate elastic shear stress. It also explains the reversibility of the transition in compressed bulk $La_3Ni_2O_7$. When pressure is released, the elastic constraint relaxes and the structure returns toward its lower-energy ambient-pressure state.

## 3. Compressed bulk nickelates

Within SSCS, the pressure threshold of bulk sample is the pressure required to suppress the ambient-pressure density-wave or related competing state and to drive a first-order structural-electronic reconstruction[1,27]. Only after this reconstruction a sufficient fraction of Ni-O-Ni bonds can enter the constrained geometry needed for superconductivity. The threshold is therefore not only a pressure scale for bond shortening, but the point at which the metastable RP lattice can access a different local symmetry and orbital state.

The same picture naturally explains why superconductivity is spatially inhomogeneous or filamentary-like in diamond-anvil-cell experiments. The distribution of local stress and strain arises from a combination of different factors, including sample shape and size, pressure medium, anisotropic compressibility, stoichiometry variation, and intergrowth phases. Only the portion of the sample falling inside the SSCS window becomes superconducting. Transport, diamagnetism, and inferred superconducting volume fraction can therefore vary strongly between samples even when the pressure environment or chemical formula are similar.

The enhancement of superconductivity by chemical substitution, hybrid RP structures, and pressure-transmitting media can be understood in the same way. Replacing La with smaller cations, constructing hybrid structures such as the 1212 phase, or using a more hydrostatic pressure medium can shift the local constraint and uniform its distribution. These routes can increase the superconducting volume fraction or change $T_c$, but they do not remove the need for an appropriate bond state which is constrained mainly by shear stress.

## 4. Epitaxial thin-film nickelates

Epitaxial films provide a complementary realization of the same physics. The substrate imposes biaxial strain, suppresses free relaxation, and can lock the film into a tetragonal-like metastable RP lattice even at ambient pressure, which produces the internal shear stress and strain. Although thin-film studies do not explicitly describe the effect as shear stress, the film is a constrained elastic object. Misfit strain, thickness-dependent relaxation, oxygen exchange and substrate-induced symmetry breaking generate internal deviatoric strain fields in favor of a superconducting configuration.

This explains why films require much lower applied pressure than bulk samples. In a film, the substrate has already supplied enough constraint needed to enter the SSCS window. Additional pressure can then tune bond lengths, the *c*-axis response, and orbital overlap to enhance $T_c$ [30,31], as illustrated in Fig.1c.

Thus, the pressure enhancement of $T_c$ in films is not the evidence that compression alone is the universal cause, rather, the compression cooperates with an already constrained shear-strained state.

## 5. Discussions and perspectives

***Implications**:* The practical implication of SSCS is a new perspective on the sample-quality sensitivity of RP nickelate superconductivity. In conventional weak-coupling materials, sample quality is primarily a disorder-scattering problem. In RP nickelates, it is also a state-selection problem. The term of SSCS emphasizes that superconductivity is selected by a constrained stress-strain state of the metastable RP lattice. Because stress is a tensor and the elastic compliance of RP nickelates is anisotropic, the relevant variable is not hydrostatic pressure alone. It is the local constrained deformation of the Ni-O framework, including the shear-like component that controls octahedral rotation, interlayer Ni-O-Ni alignment, and the coupling between the Ni $dz^2$ and $dx^2$-$y^2$ orbitals. In such a system, oxygen vacancies or interstitials locally change the Ni valence, modify the $NiO_6$ volume, and pin octahedral tilts; intergrowth defects change the thickness and registry of perovskite blocks. In films, substrate steps or roughness, misfit dislocations, thickness variation, and imperfect oxygen exchange play the analogous role. These defects do not necessarily have to directly destroy the pairing mechanism, while they can instead move local regions out of the SSCS window and interrupt the connectivity between superconducting regions.

***Explanation**:* This provides a specific explanation for why reproducibility of Ni-327 superconductivity is difficult to control. The intrinsic superconducting transition temperature of optimally constrained domains may be relatively robust, while the macroscopic onset, zero-resistance temperature, Meissner signal, and apparent superconducting volume fraction are governed by the fraction and connectivity of such domains. Sample quality therefore controls both the width of the local constraint distribution and the percolation of superconducting regions. This viewpoint connects pressure-medium sensitivity in bulk crystals and substrate-quality sensitivity in films as two manifestations of the same underlying constraint-distribution problem.

The observed shrinkage of the $NiO_2$ plane in both bulk and film samples is important,[32] but it should be viewed as a structural signature within a broader constrained state rather than as a complete explanation by itself. Two samples can have similar average *in-plane* lattice constants while differing in octahedral structure, oxygen arrangement, defect density, and local strain distribution. Those differences determine how much of a sample actually lies inside the SSCS window.

***Suggestions for further research***: The SSCS scenario suggests that reproducibility can be improved by controlling both the mean local constraint and the width of its spatial distribution. For bulk samples, useful strategies include controlled sample geometry, optimized pressure media, chemical substitutions that precondition the lattice, and synthesis protocols that reduce oxygen and intergrowth in homogeneity. At fixed nominal pressure, varying the pressure medium, sample shape, and loading geometry can tune the local shear-stress distribution, which can then be compared with spatially resolved diamagnetic response. Combining nitrogen-vacancy sensing with micro-XRD, Raman spectroscopy, and local chemical probes would enable point-by-point correlations among superconductivity, stress, stoichiometry, and octahedral distortion. For films, STEM and reciprocal-space mapping could search for the corresponding shear-like distortions through varying substrate, thickness, strain state.

In summary, the SSCS scenario provides a unified framework for understanding high-$T_c$ superconductivity in Ruddlesden-Popper nickelates across bulk and film platforms. It does not replace the roles of bond angle, bond length, orbital occupation, oxygen stoichiometry, or carrier density. Instead, it specifies the mechanical and symmetry condition under which these ingredients can cooperate to stabilize superconductivity. The apparent fragility and heterogeneity of nickelate superconductors are not extrinsic complications, but central diagnostics of the superconducting state itself. This perspective suggests concrete experimental tests and practical routes toward improved reproducibility, and connects the physics of compressed bulk crystals, epitaxial films, chemically substituted samples, and hybrid RP structures within a single conceptual framework.

**Acknowledgements**

The work was supported by the National Key Research and Development Program of China (grants 2021YFA1401800 and 2022YFA1403900).

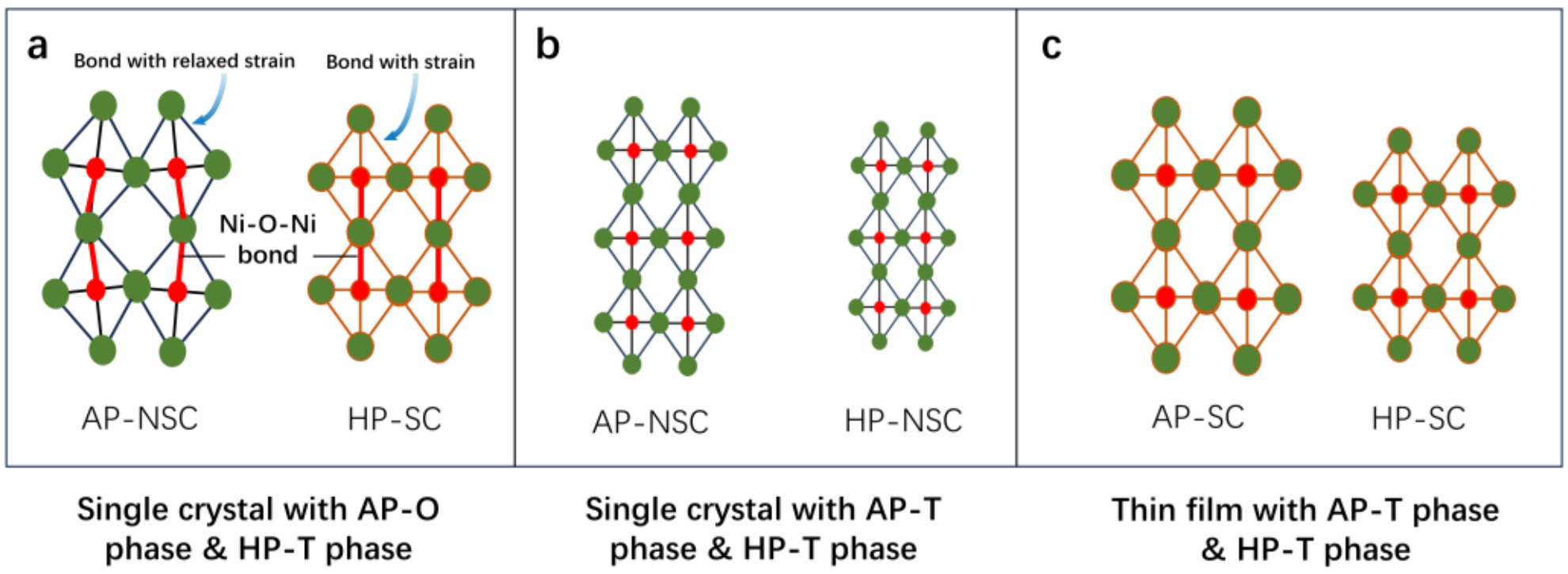

Figure 1. Schematic comparison of the RP nickelate states with and without effective elastic shear strain. (a) A compressed bulk crystal transforms from an ambient-pressure orthorhombic non-superconducting state to a high-pressure-constrained superconducting state in which selected Ni-O-Ni bonds carry elastic shear strain. (b)

An ambient-pressure tetragonal bulk phase remains non-superconducting under pressure because the required constrained shear-strain state is not produced. (c) An epitaxial thin film can be locked by the substrate into a shear-strained tetragonal-like state, and further pressure can optimize the superconducting state.